\documentclass[aps,prl,twocolumn]{revtex4-2}
\usepackage{amsmath,amssymb}
\setcitestyle{super,sort&compress}
\usepackage[colorlinks=true, citecolor=blue, linkcolor=blue, urlcolor=blue]{hyperref}
\usepackage{xcolor}
\usepackage[normalem]{ulem}

\newcommand{\blue}[1]{\textcolor{black}{#1}}
\begin{document}

\title{From Wavefunction Sign Structure to Static Correlation}

\author{Matúš Dubecký}
\email{matus.dubecky@osu.cz}
\affiliation{Department of Physics, University of Ostrava, Czech Republic}

\begin{abstract}
A variational nodal partition of the correlation energy is introduced,
\(E_{\mathrm{cor}}=E_{\mathrm{sym}}+E_{\mathrm{stat}}\),
\blue{relative to a chosen mean-field correlation baseline and its associated single-determinant node}.
Static correlation \(E_{\mathrm{stat}}\) is the \blue{negative of the} energy penalty incurred when the many-electron wavefunction is constrained to \blue{that reference node} rather than the exact one. This nodal term isolates the antisymmetric, sign-structure component of correlation, while the complementary symmetric term \(E_{\mathrm{sym}}\) necessarily contains dynamic correlation together with a distinct strong but nodeless contribution. The resulting state-based, method-independent framework clarifies the relation between dynamic, nondynamic, strong, and static correlation, places earlier node-based decompositions on rigorous footing, and explains why single-determinant fixed-node diffusion Monte Carlo can be highly accurate in some systems yet fail in others.
\end{abstract}

\maketitle

Electron correlation energy,
\begin{equation}
E_{\mathrm{cor}}=E-E_{\mathrm{MF}},
\label{eq:ecor}
\end{equation}
the missing part of the exact total energy \(E\) relative to a mean-field (MF) reference, most often Hartree-Fock (HF), is commonly partitioned into dynamic and nondynamic parts,\cite{Lowdin1959,Pople1975,London1930,Mok1996,Ramos-Cordoba2016}
\[
E_{\mathrm{cor}}=E_{\mathrm{d}}+E_{\mathrm{nd}}.
\]
Despite its wide use, this language remains ambiguous. Existing decompositions are usually tied to basis sets, orbital partitions, reference spaces, or specific methods, so neither their numerical values nor their physical interpretation are universal.\cite{Tuan1964,Cioslowski1991,Mok1996,Hollett2011,Boguslawski2012,Coe2015,Grimme2015,Ramos-Cordoba2016,Ramos-Cordoba2017,ViaNadal2019,Dunning2021,Izsak2023,Ramos-Cordoba2024,Sulka2024} The unresolved question is whether correlation can be partitioned directly from the defining exact and MF quantum states, irrespective of representation.

For real fermionic states, the natural state variable is the nodal hypersurface
\(\Gamma=\{\mathbf R:\Psi(\mathbf R)=0\}\), which partitions configuration space into path-connected constant-sign domains.\cite{Ceperley1991}
Exact and approximate nodes generally differ in both topology and shape.\cite{Mitas2006,Bressanini2012,Rasch2014} Constraining the wavefunction to an approximate node introduces a variational bias. A broad and practically important class of such approximations is provided by single-determinant (SD) nodes. When the approximate node is taken from a chosen MF SD reference, the associated bias defines a component of the correlation energy and motivates
\begin{equation}
E_{\mathrm{cor}}=E_{\mathrm{sym}}+E_{\mathrm{stat}},
\label{eq:main}
\end{equation}
where \(E_{\mathrm{stat}}\) is the \blue{negative of the} energy penalty for imposing the MF SD node instead of the exact one, and \(E_{\mathrm{sym}}\) is the complementary correlation recoverable without changing that node. The term ``static'' refers to the fixed nodal Dirichlet boundary and remains consistent with the notion of SD inadequacy. Anchoring the partition in the nodes of the states defining \(E_{\mathrm{cor}}\) makes it state- rather than method-based.

This partition immediately sharpens quantum-chemical terminology. The symmetric sector is not elementary: it contains the usual short-range dynamic correlation \(E_{\mathrm{d}}\), characteristic of closed-shell regimes, but also a distinct strong\cite{GanoeShee2024} yet nodeless contribution \(E_{\mathrm{strong}}\), characteristic of open-shell near-degeneracy. We therefore write
\begin{equation}
E_{\mathrm{sym}}=E_{\mathrm{d}}+E_{\mathrm{strong}},
\qquad
E_{\mathrm{cor}}=E_{\mathrm{d}}+E_{\mathrm{strong}}+E_{\mathrm{stat}},
\label{eq:3split}
\end{equation}
so that
\begin{equation}
E_{\mathrm{nd}}=E_{\mathrm{strong}}+E_{\mathrm{stat}}.
\label{eq:end}
\end{equation}
Thus the general split contains three terms. The decomposition of \(E_{\mathrm{sym}}\) into \(E_{\mathrm{d}}\) and \(E_{\mathrm{strong}}\) is conceptually clear but its rigorous quantification is beyond the present scope; the focus here is the nodal term \(E_{\mathrm{stat}}\).  

In practice, \(E_{\mathrm{sym}}\) can be estimated with fixed-node diffusion Monte Carlo (FNDMC)\cite{Anderson1975,Ceperley1977,Umrigar1993,Ceperley1980,Reynolds1982,Mitas1991} using the chosen MF SD node, while \(E_{\mathrm{stat}}\) then follows by difference from \(E_{\mathrm{cor}}\). The fixed-node bias is thus a well-defined part of the correlation energy systematically omitted by SD-FNDMC, explaining its uneven accuracy across systems.

\textit{Variational construction.}
Let \(E_{\mathrm{MF}}\) denote a chosen MF reference energy and \(\Gamma_\mathrm{MF}\) the node of the corresponding SD wavefunction. Equation~(\ref{eq:ecor}) defines the associated correlation energy. Using \(\Gamma\) and \(\Gamma_\mathrm{MF}\) as Dirichlet boundary conditions, define
\begin{equation}
E=\min_{\Psi\in\mathcal F[\Gamma]}\langle \Psi|\hat H|\Psi\rangle,
\qquad
E_{\mathrm{int}}=\min_{\Psi\in\mathcal F[\Gamma_\mathrm{MF}]}\langle \Psi|\hat H|\Psi\rangle,
\label{eq:var}
\end{equation}
where \(\mathcal F[\Gamma]\) denotes antisymmetric wavefunctions vanishing on the exact node \(\Gamma\), and \(\mathcal F[\Gamma_\mathrm{MF}]\) the corresponding set constrained to the MF node \blue{defining the intermediate energy \(E_{\mathrm{int}}\)}. The first variational problem is formal and simply returns \blue{the exact ground-state energy} \(E\). Its role is to define \(E_{\mathrm{stat}}\) explicitly as the difference between two variational energies determined by their underlying nodes. Since \(\Gamma_\mathrm{MF}\) is generally suboptimal, the variational principle implies
\[
E\le E_{\mathrm{int}}\le E_{\mathrm{MF}}.
\]
We then define
\begin{equation}
E_{\mathrm{sym}}\equiv E_{\mathrm{int}}-E_{\mathrm{MF}},
\qquad
E_{\mathrm{stat}}\equiv E-E_{\mathrm{int}}.
\label{eq:def}
\end{equation}
Equation~(\ref{eq:main}) follows trivially.

The meaning of Eq.~(\ref{eq:def}) is direct. \(E_{\mathrm{sym}}\) contains all correlation obtainable by amplitude relaxation at fixed SD node, whereas \(E_{\mathrm{stat}}\) requires improvement of the node itself. Because SD nodes are restricted in both shape and topology, minimization within the SD class is generally variationally suboptimal and cannot reach the exact energy. \blue{The magnitude of the resulting missing correlation is precisely the nodal penalty \(\left|E_{\mathrm{stat}}\right|\).}

The decomposition is convention dependent in numerical value, since \(E_{\mathrm{MF}}\) fixes both the correlation baseline and the reference node \(\Gamma_\mathrm{MF}\), but not in structure. Once a convention is chosen, \(E_{\mathrm{cor}}\) is fixed and the nodal split is uniquely defined. Within that convention, \(E_{\mathrm{stat}}\) has a purely variational definition and, once \(E_{\mathrm{cor}}\) and \(E_{\mathrm{sym}}\) are known, does not require explicit knowledge of \(\Gamma\) in practice.

\blue{In the numerical examples presented below, we adopt the restricted (open-shell) HF convention, \(E_{\mathrm{MF}}=E_{\mathrm{RHF}}\), which provides a clean SD baseline with a well-defined node obtained by SCF minimization within the SD space under the correct symmetry and quantum-number constraints.}

Other SD choices, such as UHF, define alternative conventions and corresponding numerical partitions,\cite{Hollett2011} \blue{whereas DFT nodes require separate care}.\cite{Per2019} The value of \(E_{\mathrm{cor}}\) is tied to the defining pair of states and their associated nodes \((\Gamma,\Gamma_{\mathrm{MF}})\). The split is meaningful only when the constraint defining \(E_{\mathrm{int}}\) uses the same reference node \(\Gamma_{\mathrm{MF}}\) that underlies \(E_{\mathrm{MF}}\), together with the exact node \(\Gamma\) entering \(E\); only then does \(E_{\mathrm{int}}\) partition \(E_{\mathrm{cor}}\) into well-defined components. If this link is broken, the split of \(E_{\mathrm{cor}}\) becomes ill-defined, whereas the variational definition of \(E_{\mathrm{stat}}\) remains \blue{well defined with respect to \(E\)}.

A DFT-based convention is generally not suitable for defining \(E_{\mathrm{cor}}\), since \(E_{\mathrm{MF}}\) already contains approximate correlation, so the top-down partition is not well-defined. Nevertheless, fixed-node calculations with such SD nodes still incur a variational bias interpretable as \blue{a nodal contribution analogous to \(E_{\mathrm{stat}}\)}, although its role as a component of \(E_{\mathrm{cor}}\) remains unclear.

Different conventions can also be used to compare the suitability of different SDs as references for post-HF treatments. Such comparisons may point toward a fully universal, convention-independent decomposition, but whether this can be achieved remains open and is beyond the present scope.


\textit{Interpretation.}
The nodal partition of Eq.~(\ref{eq:main}) reflects the \blue{conjectured} topological mismatch between exact and SD nodes: exact fermionic ground states \blue{generically} have two nodal domains, whereas generic SD wavefunctions overfragment configuration space into multiple path-connected sign domains.\cite{Mitas2006,Bressanini2012,Rasch2014}
For systems with more than \blue{two electrons}, this makes the SD nodal constraint generically suboptimal, so \(E_{\mathrm{stat}}\neq0\). A nonzero nodal penalty is thus expected beyond the simplest cases, and even nominally single-reference systems need not coincide with the SD reference. Conversely, \(E_{\mathrm{stat}}=0\) in nodeless cases or whenever the SD node is exact.

This gives \(E_{\mathrm{stat}}\) a precise many-body meaning. It is not merely ``multireference character'' in a loose orbital sense, but the energetic consequence of an imperfect fermionic sign structure defined relative to the reference node \(\Gamma_\mathrm{MF}\) associated with the chosen \(E_{\mathrm{cor}}\) convention. In this sense, \(E_{\mathrm{stat}}\) provides a state-resolved measure of fermion-sign complexity\cite{Troyer2005,Whitfield2013,Huggins2022,Choi2024} relative to the chosen MF reference.

From the conventional quantum-chemistry standpoint, a larger \(\left|E_{\mathrm{stat}}\right|\) indicates a greater departure from SD adequacy. \blue{Within a given Hamiltonian family (fixed \(N\) and interaction form, e.g., along a bond-stretching coordinate) and MF convention, a larger \(\left|E_{\mathrm{stat}}\right|\)} implies that the exact state is less well represented by the SD reference and that recovering the correct node \blue{requires} greater antisymmetric flexibility. Consequently, \blue{one expects} larger multideterminant expansions \blue{to repair the node}, and reduced-density-matrix approaches \blue{to become more sensitive} to \(N\)-representability constraints.\cite{Schlimgen2017}

\blue{The structure of Eq.~(\ref{eq:3split}) also suggests a pairing-based intuition. The symmetric sector \(E_{\mathrm{sym}}\) is the correlation accessible under a fixed SD nodal constraint and is naturally associated with singlet-like unlike-spin pairing, encompassing both closed-shell dynamic (\(E_{\mathrm{d}}\)) and open-shell near-degenerate (\(E_{\mathrm{strong}}\)) regimes. By contrast, \(E_{\mathrm{stat}}\) is the node-driven contribution: it measures the energy gained by changing the fermionic sign structure beyond the SD node and can be associated with like-spin, triplet-like pair correlations beyond the SD description.}

Singlet H\(_2\) at dissociation provides a minimal example showing that \(E_{\mathrm{sym}}\) must contain two distinct contributions: as the Coulomb interaction vanishes (one electron per atom), the remaining correlation becomes strongly nondynamic\cite{Ding2020}, yet the ground state remains nodeless. The near-degeneracy effect therefore cannot belong to \(E_{\mathrm{stat}}\) and must reside in \(E_{\mathrm{sym}}\), motivating the identification of \(E_{\mathrm{strong}}\) alongside \(E_{\mathrm{d}}\).

This resolves the standard terminology on a state basis: dynamic correlation is the short-range part of \(E_{\mathrm{sym}}\); strong correlation is the nodeless near-degeneracy part of \(E_{\mathrm{sym}}\); \blue{static correlation is the signed nodal contribution \(E_{\mathrm{stat}}\), whose magnitude is the nodal penalty;} and nondynamic correlation is the sum \(E_{\mathrm{strong}}+E_{\mathrm{stat}}\). In large-gap closed-shell systems, \(E_{\mathrm{strong}}=0\), giving
\[
E_{\mathrm{cor}}=E_{\mathrm{d}}+E_{\mathrm{stat}}.
\]

The three-component split of Eq.~(\ref{eq:3split}) provides a unified and rigorous foundation for earlier empirical classifications of correlation.\cite{Hollett2011,Coe2015} In particular, the separation of ``static''  \blue{(in the terminology of Ref.~\citenum{Hollett2011}, corresponding to nondynamic correlation here)} correlation into two flavors, one that can be captured within UHF (Type A, ``absolute near-degeneracy'') and one that cannot (Type B, ``relative near-degeneracy''), maps directly onto the present partition: the former corresponds to the strong but nodeless contribution \(E_{\mathrm{strong}}\), mimicked by spin-symmetry breaking within an SD, while the latter requires a change of nodal topology inaccessible to SD and belongs to \(E_{\mathrm{stat}}\). Likewise, the categories ``not strongly multireference and not strongly correlated,'' ``strongly multireference but not strongly correlated,'' and ``strongly multireference together with strong correlation''\cite{Coe2015} correspond to combinations of \(E_{\mathrm{d}}\), \(E_{\mathrm{strong}}\), and \(E_{\mathrm{stat}}\) that separate predominantly dynamic, strong but nodeless, and nodal regimes.

In this context, empirical post-corrections based on UHF may serve as practical proxies for \(E_{\mathrm{strong}}\), until a rigorous and universally applicable partition of \(E_{\mathrm{sym}}\) into its dynamic and strong components is developed.\cite{Sulka2023,Sulka2024}

\textit{Relation to FNDMC.}
The nodal term \(E_{\mathrm{stat}}\), defined by the sign structure of the states entering \(E_{\mathrm{cor}}\), is state-based and independent of any particular method. When evaluated with an SD MF node, it is directly reflected in the fixed-node bias of SD-FNDMC, which can be interpreted as a missing component of correlation rather than a method-specific artifact.

For an RHF node, FNDMC minimizes the energy within the constraint \(\mathcal F[\Gamma_{\mathrm{RHF}}]\) and yields \(E_{\mathrm{int}}\) of Eq.~(\ref{eq:var}) in the exact projection limit. SD-FNDMC therefore recovers \(E_{\mathrm{sym}}\) but omits \(E_{\mathrm{stat}}\) [Eq.~(\ref{eq:def})]:
\[
E_{\mathrm{stat}}=E-E_{\mathrm{int}}\approx E-E_{\mathrm{DMC}}[\Gamma_{\mathrm{RHF}}].
\]
\blue{The fixed-node bias in this context is therefore a precisely defined missing part of \(E_{\mathrm{cor}}\).} Energy differences obtained with a common nodal ansatz therefore depend on cancellation of differences in \(E_{\mathrm{stat}}\), rather than on elimination of an uncontrolled error. Because SD nodal surfaces converge rapidly with one-particle basis quality,\cite{Dubecky2017jctc} the corresponding estimator is numerically robust.

The same interpretation applies to alternative SD nodes, including those from DFT.\cite{Per2019} Although a DFT-based definition of \(E_{\mathrm{cor}}\) is ambiguous, since the nodal shape is determined in the presence of approximate correlation and may mix correlation effects that obscure a clean partition, the fixed-node energy still defines a variational penalty relative to the exact state. The residual fixed-node error therefore reflects \blue{a quantity analogous to} \(E_{\mathrm{stat}}\), and energy differences again rely on its cancellation.

This perspective thus resolves the SD-FNDMC puzzle by identifying the fixed-node bias as a physical component of the correlation hierarchy. With a fixed SD node, SD-FNDMC recovers \(E_{\mathrm{sym}}\) but omits \(E_{\mathrm{stat}}\), so its accuracy in energy differences is controlled by the magnitude and cancellation of \(E_{\mathrm{stat}}\). It performs well when correlation is dominated by the symmetric sector, including strong but nodeless near-degeneracy effects such as in correlated solids (e.g., FeO, VO$_2$),\cite{Kolorenc2008,Zheng2015} but can fail when small, noncanceling nodal contributions set the relevant energy scale, as in the benzene dimer.\cite{Fanta2025} High accuracy may arise from cancellation of \(E_{\mathrm{stat}}\),\cite{Dubecky2017} but this is not guaranteed \textit{a priori}. Within this framework, SD-FNDMC is best viewed as a stochastic solver for \(E_{\mathrm{sym}}\), with \(E_{\mathrm{stat}}\) obtained by complementarity.

The same perspective also clarifies why single-reference configuration-based methods and SD-FNDMC can behave differently.\cite{Benali2020,AlHamdani2021,Fanta2025,Gruneis2025NatComm} In regimes where the leading nodal correction can be captured by low-order excitations from the SD reference, methods such as CI or coupled cluster can inherently repair the SD node, whereas SD-FNDMC, by construction, cannot. In the present language, the former can recover the leading part of \(E_{\mathrm{stat}}\), subject to basis and expansion limitations, while the latter remains confined to \(E_{\mathrm{sym}}\) within a fixed SD node, leading to uneven biases that propagate to energy differences. Practical \textit{a priori} estimators of \(E_{\mathrm{stat}}\), for example minimal configuration models that reproduce the correct nodal topology,\cite{Rasch2012,Kulahlioglu2014} are therefore valuable for diagnosing potentially dangerous SD fixed-node bias.

This perspective also suggests a constructive role for beyond-SD trial states. Compact topology-aware ansätze, including pairing forms,\cite{Bajdich2006,Sorella2007} as well as neural or hybrid antisymmetric-symmetric architectures,\cite{Hermann2020,Pfau2020} can be viewed as approaches to recover \(E_{\mathrm{stat}}\) by improving the antisymmetric sector while retaining efficient descriptions of the symmetric sector.

\textit{Numerical examples.}
The limiting cases are immediate. For two-electron singlet ground states such as He, or more generally for any nodeless state, \(E_{\mathrm{stat}}=0\) and all correlation belongs to \(E_{\mathrm{sym}}\). In the lowest triplet state of He, the exact node is known and coincides with the SD node,\cite{Bressanini2005,Datta2020} so again \(E_{\mathrm{stat}}=0\). Assuming no strong symmetric component, one recovers \(E_{\mathrm{cor}}=E_{\mathrm{d}}\).

Consistent with the topological picture above, \(E_{\mathrm{stat}}\) is generally nonzero \blue{in many-electron systems once the SD node no longer matches the exact one}. Table~\ref{tab:nodal_partition} reports proof-of-principle RHF-based nodal partitions of the valence correlation energy for selected second-period atoms and diatomics. The data show a clear increase of \(\left|E_{\mathrm{stat}}\right|\) with electron number and near-degeneracy, while \(\left|E_{\mathrm{sym}}\right|\) remains the dominant contribution across the cases considered. The ratio \(\left|E_{\mathrm{stat}}\right|/\left|E_{\mathrm{cor}}\right|\) is small for closed-shell atoms but becomes larger for open-shell configurations and low-lying excited states, indicating progressive failure of the SD node. In particular, O \(^1\!D\) exhibits a substantially larger \(\left|E_{\mathrm{stat}}\right|\) than O \(^3\!P\), consistent with its stronger multiconfigurational character.

For molecules, bond stretching amplifies both \(\left|E_{\mathrm{cor}}\right|\) and \(\left|E_{\mathrm{stat}}\right|\), but with distinct trends. In BH, the increase is smooth and reflects the gradual onset of static correlation. In F\(_2\), the growth of \(\left|E_{\mathrm{stat}}\right|\) is steeper and accompanied by nonmonotonic behavior near dissociation, indicating a nonuniform deterioration of the SD nodal description. Overall, the table shows that \(\left|E_{\mathrm{stat}}\right|\) follows the usual quantum-chemical notion of static correlation and provides a quantitative measure of SD nodal inadequacy, motivating application to a broader range of systems.

\begin{table}[t]
\caption{Nodal partition of valence correlation energy \blue{magnitudes} for selected second-period atoms and diatomics. \(N\) is the number of valence electrons. Data adapted from Ref.~\citenum{Sulka2023}.}
\label{tab:nodal_partition}
\begin{ruledtabular}
\begin{tabular}{lccrrrr}
System & \(N\) & State & \(\left|E_{\rm cor}\right|\) & \(\left|E_{\rm sym}\right|\) & \(\left|E_{\rm stat}\right|\) & \(\left|E_{\rm stat}\right|/\left|E_{\rm cor}\right|\) \\
       &       &       & (mHa)           & (mHa)            & (mHa)            & (\%)                        \\
\hline
O       & 6  & \(^3\!P\)  & 194.85 & 180.59 & 14.26 &  7.3 \\
        &    & \(^1\!D\)  & 236.32 & 197.88 & 38.44 & 16.3 \\
F       & 7  & \(^2\!P\)  & 259.64 & 245.42 & 14.22 &  5.5 \\
Ne      & 8  & \(^1\!S\)  & 332.98 & 314.06 & 18.92 &  5.7 \\
\hline
BH      & 4  & 1.23~\AA{} & 108.13 &  95.93 & 12.20 & 11.3 \\
        &    & 2.00~\AA{} & 139.10 & 123.70 & 15.40 & 11.1 \\
        &    & 3.00~\AA{} & 164.42 & 148.37 & 16.05 &  9.8 \\
\hline
F\(_2\) & 14 & 1.30~\AA{} & 611.55 & 573.72 & 37.83 &  6.2 \\
        &    & 2.00~\AA{} & 721.84 & 663.59 & 58.25 &  8.1 \\
        &    & 2.80~\AA{} & 828.65 & 778.58 & 50.07 &  6.0 \\
\end{tabular}
\end{ruledtabular}
\end{table}

\textit{Summary.}
\blue{A variational nodal partition of the correlation energy was introduced that defines static correlation as the signed contribution associated with the penalty for imposing the MF SD node rather than the exact one.} The decomposition
\[
E_{\mathrm{cor}}=E_{\mathrm{sym}}+E_{\mathrm{stat}}
\]
isolates a sign-structure contribution \(E_{\mathrm{stat}}\) and separates it from the symmetric amplitude contribution \(E_{\mathrm{sym}}\). Since \(E_{\mathrm{sym}}\) contains both dynamic and strong but nodeless contributions, the general correlation split is unavoidably three-term,
\[
E_{\mathrm{cor}}=E_{\mathrm{d}}+E_{\mathrm{strong}}+E_{\mathrm{stat}},
\qquad
E_{\mathrm{nd}}=E_{\mathrm{strong}}+E_{\mathrm{stat}}.
\]
This establishes a uniquely defined, variationally exact, state-based decomposition within a fixed MF convention and clarifies the relation between dynamic, strong, nondynamic, and static correlation. The framework also recasts the SD fixed-node error of DMC as a missing, physically identifiable component of the correlation energy, unifying the correlation terminology, nodal structure, and fixed-node bias within a single coherent variational framework.

\begin{acknowledgments}
We thank D.~Bressanini, \blue{M.~Ditte}, K.~D.~Jordan, P.~Jurečka, M.~Šulka, and M.~Novotný for insightful discussions. Financial support from EU LERCO project CZ.10.03.01/00/22\_003/0000003 is gratefully acknowledged.
\end{acknowledgments}

\bibliography{references.bib}

@article{Lowdin1959,
  author  = {L{\"o}wdin, P.-O.},
  title   = {Correlation Problem in Many-Electron Quantum Mechanics I. Review of Different Approaches and Discussion of Some Current Ideas},
  journal = {Adv. Chem. Phys.},
  year    = {1959},
  volume  = {2},
  pages   = {207-322},
}

@article{Cioslowski1991,
  author  = {Cioslowski, J.},
  title   = {Density-Driven Self-Consistent-Field Method: Density-Constrained Correlation Energies in the Helium Series},
  journal = {Phys. Rev. A},
  year    = {1991},
  volume  = {43},
  pages   = {1223-1228},
  doi     = {10.1103/PhysRevA.43.1223},
}

@article{Mok1996,
  author  = {Mok, K. W. and Neumann, R. and Handy, N. C.},
  title   = {Dynamical and Nondynamical Correlation},
  journal = {J. Phys. Chem.},
  year    = {1996},
  volume  = {100},
  pages   = {6225-6230},
  doi     = {10.1021/jp9528020},
}

@article{Ramos-Cordoba2016,
  author  = {Ramos-C{\'o}rdoba, E. and Salvador, P. and Matito, E.},
  title   = {Separation of Dynamic and Nondynamic Correlation},
  journal = {Phys. Chem. Chem. Phys.},
  year    = {2016},
  volume  = {18},
  pages   = {24015-24023},
  doi     = {10.1039/C6CP03072F},
}

@article{Anderson1975,
  author  = {Anderson, J. B.},
  title   = {A Random-Walk Simulation of the Schr{\"o}dinger Equation: H$_3^+$},
  journal = {J. Chem. Phys.},
  year    = {1975},
  volume  = {63},
  pages   = {1499-1502},
  doi     = {10.1063/1.431514},
}

@article{Reynolds1982,
  author  = {Reynolds, P. J. and Ceperley, D. M. and Alder, B. J. and Lester, W. A.},
  title   = {Fixed-Node Quantum Monte Carlo for Molecules},
  journal = {J. Chem. Phys.},
  year    = {1982},
  volume  = {77},
  pages   = {5593-5603},
  doi     = {10.1063/1.443766},
}

@article{Umrigar1993,
  author  = {Umrigar, C. J. and Nightingale, M. P. and Runge, K. J.},
  title   = {A Diffusion Monte Carlo Algorithm with Very Small Time-Step Errors},
  journal = {J. Chem. Phys.},
  year    = {1993},
  volume  = {99},
  pages   = {2865-2890},
  doi     = {10.1063/1.465195},
}

@article{Rasch2012,
  author  = {Rasch, K. M. and Mitas, L.},
  title   = {Impact of Electron Density on the Fixed-Node Errors in Quantum Monte Carlo of Atomic Systems},
  journal = {Chem. Phys. Lett.},
  year    = {2012},
  volume  = {528},
  pages   = {59-62},
  doi     = {10.1016/j.cplett.2012.01.016},
}

@article{Kulahlioglu2014,
  author  = {Kulahlioglu, A. H. and Rasch, K. M. and Hu, S. and Mitas, L.},
  title   = {Density Dependence of Fixed-Node Errors in Diffusion Quantum Monte Carlo: Triplet Pair Correlations},
  journal = {Chem. Phys. Lett.},
  year    = {2014},
  volume  = {591},
  pages   = {170-174},
  doi     = {10.1016/j.cplett.2013.11.033},
}

@article{Bajdich2006,
  author  = {Bajdich, M. and Mitas, L. and Drobny, G. and Wagner, L. K. and Schmidt, K. E.},
  title   = {Pfaffian Pairing Wave Functions in Electronic-Structure Quantum Monte Carlo},
  journal = {Phys. Rev. Lett.},
  year    = {2006},
  volume  = {96},
  pages   = {130201},
}

@article{Dubecky2017,
  author  = {Dubeck{\'y}, M.},
  title   = {Bias Cancellation in One-Determinant Fixed-Node Diffusion Monte Carlo: Insights from Fermionic Occupation Numbers},
  journal = {Phys. Rev. E},
  year    = {2017},
  volume  = {95},
  number  = {3},
  pages   = {033308},
  doi     = {10.1103/PhysRevE.95.033308},
}

@article{Coe2015,
  author  = {Coe, J. P. and Paterson, M. J.},
  title   = {Investigating Multireference Character and Correlation in Quantum Chemistry},
  journal = {J. Chem. Theory Comput.},
  year    = {2015},
  volume  = {11},
  pages   = {4189-4196},
}

@article{Pople1975,
  author  = {Pople, J. A. and Binkley, J. S.},
  title   = {Correlation Energies for AH$_n$ Molecules and Cations},
  journal = {Mol. Phys.},
  year    = {1975},
  volume  = {29},
  pages   = {599-611},
  doi     = {10.1080/00268977500100511},
}

@article{Hollett2011,
  author  = {Hollett, J. W. and Gill, P. M. W.},
  title   = {The Two Faces of Static Correlation},
  journal = {J. Chem. Phys.},
  year    = {2011},
  volume  = {134},
  pages   = {114111},
  doi     = {10.1063/1.3570574},
}

@article{London1930,
  author  = {London, F.},
  title   = {Zur Theorie und Systematik der Molekularkr{\"a}fte},
  journal = {Z. Phys.},
  year    = {1930},
  volume  = {63},
  pages   = {245-279},
}

@article{Mitas1991,
  author  = {Mitas, L. and Shirley, E. L. and Ceperley, D. M.},
  title   = {Nonlocal Pseudopotentials and Diffusion Monte Carlo},
  journal = {J. Chem. Phys.},
  year    = {1991},
  volume  = {95},
  pages   = {3467-3475},
  doi     = {10.1063/1.460849},
}

@article{Ceperley1977,
  author  = {Ceperley, D. M. and Chester, G. V. and Kalos, M. H.},
  title   = {Monte Carlo Simulation of a Many-Fermion Study},
  journal = {Phys. Rev. B},
  year    = {1977},
  volume  = {16},
  pages   = {3081-3099},
  doi     = {10.1103/PhysRevB.16.3081},
}

@article{Ding2020,
  author  = {Ding, L. and Schilling, C.},
  title   = {Correlation Paradox of the Dissociation Limit: A Quantum Information Perspective},
  journal = {J. Chem. Theory Comput.},
  year    = {2020},
  volume  = {16},
  number  = {7},
  pages   = {4159-4175},
  doi     = {10.1021/acs.jctc.0c00054},
}

@article{Dubecky2017jctc,
  author  = {Dubeck{\'y}, M.},
  title   = {Noncovalent Interactions by Fixed-Node Diffusion Monte Carlo: Convergence of Nodes and Energy Differences vs Gaussian Basis-Set Size},
  journal = {J. Chem. Theory Comput.},
  year    = {2017},
  volume  = {13},
  number  = {8},
  pages   = {3626-3635},
  doi     = {10.1021/acs.jctc.7b00537},
}

@article{Kolorenc2008,
  author  = {Koloren{\v c}, J. and Mitas, L.},
  title   = {Quantum Monte Carlo Calculations of Structural Properties of FeO Under Pressure},
  journal = {Phys. Rev. Lett.},
  year    = {2008},
  volume  = {101},
  number  = {18},
  pages   = {185502},
  doi     = {10.1103/PhysRevLett.101.185502},
}

@article{Zheng2015,
  author  = {Zheng, H. and Wagner, L. K.},
  title   = {Computation of the Correlated Metal-Insulator Transition in Vanadium Dioxide from First Principles},
  journal = {Phys. Rev. Lett.},
  year    = {2015},
  volume  = {114},
  number  = {17},
  pages   = {176401},
  doi     = {10.1103/PhysRevLett.114.176401},
}

@article{ViaNadal2019,
  author  = {Via-Nadal, M. and Rodr{\'\i}guez-Mayorga, M. and Ramos-C{\'o}rdoba, E. and Matito, E.},
  title   = {Singling Out Dynamic and Nondynamic Correlation},
  journal = {J. Phys. Chem. Lett.},
  year    = {2019},
  volume  = {10},
  number  = {14},
  pages   = {4032-4037},
  doi     = {10.1021/acs.jpclett.9b01376},
}

@article{Ramos-Cordoba2017,
  author  = {Ramos-C{\'o}rdoba, E. and Matito, E.},
  title   = {Local Descriptors of Dynamic and Nondynamic Correlation},
  journal = {J. Chem. Theory Comput.},
  year    = {2017},
  volume  = {13},
  number  = {6},
  pages   = {2705-2711},
  doi     = {10.1021/acs.jctc.7b00293},
}

@article{Sulka2023,
  author  = {{\v S}ulka, M. and {\v S}ulkov{\'a}, K. and Jure{\v c}ka, P. and Dubeck{\'y}, M.},
  title   = {Dynamic and Nondynamic Electron Correlation Energy Decomposition Based on the Node of the Hartree-Fock Slater Determinant},
  journal = {J. Chem. Theory Comput.},
  year    = {2023},
  volume  = {19},
  number  = {22},
  pages   = {8147-8155},
  doi     = {10.1021/acs.jctc.3c00828},
}

@article{Mitas2006,
  author  = {Mitas, L.},
  title   = {Structure of Fermion Nodes and Nodal Cells},
  journal = {Phys. Rev. Lett.},
  volume  = {96},
  number  = {24},
  pages   = {240402},
  year    = {2006},
  doi     = {10.1103/PhysRevLett.96.240402},
}

@article{Ramos-Cordoba2024,
  author  = {Xu, X. and Soriano-Agueda, L. and L{\'o}pez, X. and Ramos-C{\'o}rdoba, E. and Matito, E.},
  title   = {All-Purpose Measure of Electron Correlation for Multireference Diagnostics},
  journal = {J. Chem. Theory Comput.},
  year    = {2024},
  volume  = {20},
  number  = {2},
  pages   = {721-727},
}

@article{Ceperley1991,
  author  = {Ceperley, D. M.},
  title   = {Fermion Nodes},
  journal = {J. Stat. Phys.},
  year    = {1991},
  volume  = {63},
  pages   = {1237-1267},
  doi     = {10.1007/BF01030009},
}

@article{Bressanini2012,
  author  = {Bressanini, D.},
  title   = {Implications of the Two Nodal Domains Conjecture for Ground-State Fermionic Wave Functions},
  journal = {Phys. Rev. B},
  year    = {2012},
  volume  = {86},
  number  = {11},
  pages   = {115120},
  doi     = {10.1103/PhysRevB.86.115120},
}

@article{GanoeShee2024,
  author  = {Ganoe, B. and Shee, J.},
  title   = {On the Notion of Strong Correlation in Electronic Structure Theory},
  journal = {Faraday Discuss.},
  year    = {2024},
  volume  = {254},
  pages   = {53-75},
  doi     = {10.1039/D4FD00066H},
}

@article{Bressanini2005,
  author  = {Bressanini, D. and Reynolds, P. J.},
  title   = {Unexpected Symmetry in the Nodal Structure of the He Atom},
  journal = {Phys. Rev. Lett.},
  year    = {2005},
  volume  = {95},
  number  = {11},
  pages   = {110201},
  doi     = {10.1103/PhysRevLett.95.110201},
}

@article{Tuan1964,
  author    = {Tuan, D. F. and Sinano{\u g}lu, O.},
  title     = {Many-Electron Theory of Atoms and Molecules. IV. Be Atom and Its Ions},
  journal   = {J. Chem. Phys.},
  year      = {1964},
  volume    = {41},
  number    = {9},
  pages     = {2677-2688},
  doi       = {10.1063/1.1726173},
}

@article{Rasch2014,
  author       = {Rasch, K. M. and Hu, S. and Mitas, L.},
  title        = {Communication: Fixed-Node Errors in Quantum Monte Carlo: Interplay of Electron Density and Node Nonlinearities},
  journal      = {J. Chem. Phys.},
  volume       = {140},
  pages        = {041102},
  year         = {2014},
  doi          = {10.1063/1.4862496},
}

@article{Boguslawski2012,
  author  = {Boguslawski, K. and Tecmer, P. and Legeza, {\"O}. and Reiher, M.},
  title   = {Entanglement Measures for Single- and Multireference Correlation Effects},
  journal = {J. Phys. Chem. Lett.},
  volume  = {3},
  number  = {21},
  pages   = {3129-3135},
  year    = {2012},
  doi     = {10.1021/jz301191m},
}

@article{Sulka2024,
  author  = {{\v S}ulka, M. and {\v S}ulkov{\'a}, K. and Dubeck{\'y}, M.},
  title   = {Unveiling Hidden Dynamic Correlations in CASSCF Correlation Energies by Hartree-Fock Nodes},
  journal = {J. Chem. Phys.},
  volume  = {161},
  number  = {11},
  pages   = {114112},
  year    = {2024},
  doi     = {10.1063/5.0223733},
}

@article{Datta2020,
  author  = {Datta, S. and Rejcek, J. M.},
  title   = {Nodal Structures of Few Electron Atoms},
  journal = {Eur. Phys. J. Plus},
  volume  = {135},
  number  = {2},
  pages   = {254},
  year    = {2020},
  doi     = {10.1140/epjp/s13360-020-00134-z},
}

@article{Izsak2023,
  author  = {Izs{\'a}k, R. and Ivanov, A. V. and Blunt, N. S. and Holzmann, N. and Neese, F.},
  title   = {Measuring Electron Correlation: The Impact of Symmetry and Orbital Transformations},
  journal = {J. Chem. Theory Comput.},
  year    = {2023},
  volume  = {19},
  number  = {10},
  pages   = {2703-2720},
  doi     = {10.1021/acs.jctc.3c00122},
}

@article{Per2019,
  author  = {Per, M. C. and Fletcher, E. K. and Cleland, D. M.},
  title   = {Density Functional Orbitals in Quantum Monte Carlo: The Importance of Accurate Densities},
  journal = {J. Chem. Phys.},
  year    = {2019},
  volume  = {150},
  number  = {18},
  pages   = {184101},
  doi     = {10.1063/1.5080744},
}

@article{Hermann2020,
  author  = {Hermann, Jan and Sch{\"a}tzle, Zeno and No{\'e}, Frank},
  title   = {Deep Neural Network Solution of the Electronic Schr{\"o}dinger Equation},
  journal = {Nat. Chem.},
  year    = {2020},
  volume  = {12},
  number  = {10},
  pages   = {891-897},
  doi     = {10.1038/s41557-020-0544-y},
}

@article{Pfau2020,
  author  = {Pfau, David and Spencer, George and Matthews, Adam G. D. G. and Fuchs, George},
  title   = {Ab-Initio Solution of the Many-Electron Schr{\"o}dinger Equation with Deep Neural Networks},
  journal = {Phys. Rev. Research},
  year    = {2020},
  volume  = {2},
  number  = {3},
  pages   = {033429},
  doi     = {10.1103/PhysRevResearch.2.033429},
}

@article{Dunning2021,
  author  = {Dunning, T. H., Jr. and Xu, L.-T. and Cooper, D. L. and Karadakov, P. B.},
  title   = {Spin-Coupled Generalized Valence Bond Theory: New Perspectives on the Electronic Structure of Molecules and Chemical Bonds},
  journal = {J. Phys. Chem. A},
  year    = {2021},
  volume  = {125},
  pages   = {2021-2050},
  doi     = {10.1021/acs.jpca.0c11265},
}

@article{Whitfield2013,
  author    = {Whitfield, J. D. and Love, P. J. and Aspuru-Guzik, A.},
  title     = {Computational Complexity in Electronic Structure},
  journal   = {Phys. Chem. Chem. Phys.},
  volume    = {15},
  pages     = {397-411},
  year      = {2013},
  doi       = {10.1039/C2CP42695A},
}

@article{Choi2024,
  author    = {Choi, S. and Loaiza, I. and Lang, R. A. and Mart{\'\i}nez-Mart{\'\i}nez, L. A. and Izmaylov, A. F.},
  title     = {Probing Quantum Efficiency: Exploring System Hardness in Electronic Ground State Energy Estimation},
  journal   = {J. Chem. Theory Comput.},
  year      = {2024},
  volume    = {20},
  number    = {14},
  pages     = {5982-5993},
  doi       = {10.1021/acs.jctc.4c00298},
}

@article{Troyer2005,
  author    = {Troyer, M. and Wiese, U.-J.},
  title     = {Computational Complexity and Fundamental Limitations to Fermionic Quantum Monte Carlo Simulations},
  journal   = {Phys. Rev. Lett.},
  volume    = {94},
  number    = {17},
  pages     = {170201},
  year      = {2005},
  doi       = {10.1103/PhysRevLett.94.170201},
}

@article{Grimme2015,
  author  = {Grimme, S. and Hansen, A.},
  title   = {A Practicable Real-Space Measure and Visualization of Static Electron-Correlation Effects},
  journal = {Angew. Chem. Int. Ed.},
  year    = {2015},
  volume  = {54},
  number  = {42},
  pages   = {12308-12313},
  doi     = {10.1002/anie.201501887},
}

@article{Ceperley1980,
  author  = {Ceperley, D. M. and Alder, B. J.},
  title   = {Ground State of the Electron Gas by a Stochastic Method},
  journal = {Phys. Rev. Lett.},
  volume  = {45},
  pages   = {566},
  year    = {1980},
  doi     = {10.1103/PhysRevLett.45.566},
}

@article{Huggins2022,
  author = {Huggins, W. J. and O'Gorman, B. A. and Rubin, N. C. and Reichman, D. R. and Babbush, R. and Lee, J.},
  title  = {Unbiasing Fermionic Quantum Monte Carlo with a Quantum Computer},
  journal = {Nature},
  volume = {603},
  pages  = {416-420},
  year   = {2022},
  doi    = {10.1038/s41586-021-04351-z},
}

@article{Schlimgen2017,
  author = {Schlimgen, Anthony W. and Mazziotti, David A.},
  title  = {Static and Dynamic Electron Correlation in the Ligand Noninnocent Oxidation of Nickel Dithiolates},
  journal = {J. Phys. Chem. A},
  volume = {121},
  number = {48},
  pages  = {9377-9384},
  year   = {2017},
  doi    = {10.1021/acs.jpca.7b09567},
}

@article{Fanta2025,
  author  = {Fanta, R. and Jure{\v c}ka, P. and Dubeck{\'y}, M.},
  title   = {Why Nondynamic Correlation Matters for $\pi\pi$ Stacking? Lessons from the Benzene Dimer},
  journal = {J. Phys. Chem. Lett.},
  volume  = {16},
  number  = {42},
  pages   = {10982--10988},
  year    = {2025},
  doi     = {10.1021/acs.jpclett.5c02576},
}

@article{Benali2020,
  author  = {Benali, A. and Shin, H. and Heinonen, O.},
  title   = {Quantum Monte Carlo Benchmarking of Large Noncovalent Complexes in the L7 Benchmark Set},
  journal = {J. Chem. Phys.},
  year    = {2020},
  volume  = {153},
  number  = {19},
  pages   = {194113},
  doi     = {10.1063/5.0026275},
}

@article{AlHamdani2021,
  author  = {Al-Hamdani, Y. S. and Nagy, P. R. and Zen, A. and Barton, D. and K{\'a}llay, M. and Brandenburg, J. G. and Tkatchenko, A.},
  title   = {Interactions between Large Molecules Pose a Puzzle for Reference Quantum Mechanical Methods},
  journal = {Nat. Commun.},
  year    = {2021},
  volume  = {12},
  pages   = {3927},
  doi     = {10.1038/s41467-021-24119-3},
}

@article{Sorella2007,
  author  = {Sorella, S. and Casula, M. and Rocca, D.},
  title   = {Weak Binding between Two Aromatic Rings: Feeling the van der Waals Attraction by Quantum Monte Carlo Methods},
  journal = {J. Chem. Phys.},
  year    = {2007},
  volume  = {127},
  number  = {1},
  pages   = {014105},
  doi     = {10.1063/1.2746035},
}

@article{Gruneis2025NatComm,
  author  = {Sch{\"a}fer, T. and Irmler, A. and Gallo, A. and Gr{\"u}neis, A.},
  title   = {Understanding Discrepancies in Noncovalent Interaction Energies from Wavefunction Theories for Large Molecules},
  journal = {Nat. Commun.},
  year    = {2025},
  volume  = {16},
  pages   = {9108},
  doi     = {10.1038/s41467-025-64104-8},
}
\end{document}